\documentstyle[12pt]{article}
\newcommand{\be}{\begin{equation}}
\newcommand{\ee}{\end{equation}}
\newcommand{\ba}{\begin{eqnarray}}
\newcommand{\ea}{\end{eqnarray}}

\begin{document}
\bigskip
\begin{center}
{\bf\Large
Hamilton-Jacobi treatment of a non-relativistic particle \\
         on a curved space}
\end{center}
\bigskip

\bigskip
\begin{center}
{\bf Dumitru Baleanu}\footnote[1]{On leave of absence from
Institute of Space Sciences, P.O.BOX, MG-23, R
76900, Magurele-Bucharest, Romania,\, E-mail: baleanu@venus.nipne.ro } and
{\bf Yurdahan G\"{u}ler}\footnote[2]{E-Mail:~~yurdahan@cankaya.edu.tr}
\end{center}
\begin{center}
Department of Mathematics and Computer Sciences, Faculty of Arts
and Sciences, Cankaya University-06530, Ankara , Turkey
\end{center}
\begin{abstract}
In this paper a non-relativistic particle moving on a hypersurface in a
curved space and the multidimensional rotator
are investigated using the Hamilton-Jacobi formalism.
The equivalence with the Dirac Hamiltonian formalism is
demonstrated in both Cartesian and curvilinear coordinates.
The energy spectrum of the multidimensional rotator is equal to that
of a pure Laplace-Beltrami operator with no additional constant
arising from the curvature of the sphere.
 \end{abstract}
\newpage
\section {Introduction}

Constraint Hamiltonian systems play a crucial role in gauge theories.
Since Dirac's pioneering work \cite{Dirac} on constrained systems, there has been considerable
progress in this field \cite{A, K, H, SH}.
Although some basic steps were taken, there are still some more problems which
 need deeper analysis. Especially one should define the Dirac brackets explicitly to quantize second-class
constraints. But this is not an easy task, because except for very particular cases,
e.g. the Dirac brackets are c-numbers, this problem does not have a general solution.
In other words, it is extremely difficult to find a representation for the
independent operators.Determination of degre of freedom of a singular
system is a  vital problem especially if  second- class constraints
exist. In fact, the reduced  phase space is a symplectic manifold in
 mathematical language
\cite{w} and the Darboux theorem ensures that one can find, at least
locally, the coordinates
in terms of which the Poisson brackets (defined on the reduced phase space
in the presence of constraints) have the canonical form.

Quantization of a free point particle in curved space is a
long-standing and controversial problem in quantum
mechanics \cite{witt, marinov}. Dirac
has emphasized that canonical quantization rules are consistent
only in a Cartesian reference frame.
Attempts to generalize these
rules to curved space run into the notorious operator-ordering
problem of momentum and coordinates \cite{Dirac}. Podolski avoided
this problem \cite{podolski} by postulating that the Laplacian in the free
Schr\"{o}dinger operator $H=-{{\hbar}^2\bigtriangleup\over
2}$ should be replaced by the Laplace-Beltrami operator
$\bigtriangleup_{LB}=g^{-1\over 2}{\partial_{\mu}}g^{1\over
2}g^{\mu\nu}{\partial_{\nu}}$, where
$\partial_{\mu}={\partial\over\partial q^{\mu}}$ are partial derivatives
with respect to the N- dimensional curved space coordinates,
and g is the determinant of the metric $g_{\mu\nu}(q)$.This postulate has
generally been accepted as being correct since it yields , for a
N-dimensional Cartesian space with coordinates  $x^{i}$, an energy
${{\hat L}^{2}_{a}\over 2R^2}$.Here ${\hat L_{a}}=-i {\hat
p}_{i}(L_{a})_{ij}x_{j}$ with ${\hat
p}_{i}=-i{\hbar}{\partial\over\partial x_{i}}$ are the unique
quantum-mechanical differential operator representation of the
${N(N+1)\over 2}$ generators $L_{a}$ of the rotation group
$SO(N+1)$ in flat space \cite{kleinert}.
A discrepancy with the Dirac formalism has however  been reported  in
several papers \cite{marinov, kleinert, girotti},
as well as other different results \cite{omote, saa}.
In spite of all these developments, the status of the problem
is very confusing, and there have been many papers claiming
a rejection of the Dirac formalism \cite{kleinert}, an intrinsic difference
between path formulation and operator formalism \cite{girotti},
or advocating different quantization schemes \cite{omote}.
Besides, in the definition of the Wheeler-de-Witt equation
\cite{bertolami}, which has the central importance in string theories,
in curved space this problem arises again.

An alternative method of quantization is the Hamilton-Jacobi
formulation initiated by one of us \cite{gu5, gu6, gu7}.
Using the Carath\'{e}odory's equivalent Lagrangian
method  we find a set of Hamilton-Jacobi equations integrated by the method of
characteristics \cite{caratheo, kastrup}.
  Recently this formalism was generalized to the singular systems
with higher order Lagrangians and to systems which have  elements
of the Berezin algebra \cite{p11, p12, p13}.
Even more recently the quantization of
the systems with constraints was investigated using this approach
\cite{gu14, gu15, gu16}. The  advantage  of using the Hamilton-Jacobi formalism
 is that we have no difference between first and second class
constraints and we do not need gauge fixing term because
the gauge variables are separated in the processs of constructing  an
integrable system of total differential equations.
In addition the action provided by the formalism can be
used in the process of path integral quantization method of the constrained systems.
However, the quantization of the systems with second class constraints is
problematic for Hamilton-Jacobi formalism because the system of
equations is not integrable. To solve this problem we have
two basic possibilities ,the first one is  to enlarge the phase
space \cite{fad} and the other one is to keep to the original phase space
itself \cite{mitra, gar}.

Let us consider a  singular Lagrangian with Hessian
matrix of rank n-r . The formalism leads us to the following
Hamiltonians

\be\label{doi} H_{\alpha}^{'}=H_{\alpha}(t_{\beta}, q_{a}, p_{a})
+p_{\alpha}, \ee where $\alpha,\beta=n-r +1,\cdots, n$, $a=1, \cdots
n-r$. The usual hamiltonian $H_0$ is defined as

\be\label{unu} H_{0}=-L(t, q_{i}, {\dot q_{\nu}}, {\dot q_{a}=w_{a}})
+p_{a}w_{a} + {\dot
q_{\mu}}p_{\mu}\mid_{p_{\nu}=-H_{\nu}},\nu=0, n-r+1, \cdots, n. \ee
which is independent of $\dot q_{\mu}$. Here ${\displaystyle \dot
q_{a}={dq_{a}\over d\tau}}$, where $\tau$ is a parameter and $\omega_{a}$
are obtained from the definition of generalized momenta.
 The equations of motion are obtained as total differential equations
in many variables as follows

\be\label{{pq}} dq_{a}={\partial H_{\alpha}^{'}\over\partial
p_{a}}dt_{\alpha}, dp_{a}=-{\partial H_{\alpha}^{'}\over\partial
q_{a}}dt_{\alpha}, dp_{\mu}=-{\partial H_{\alpha}^{'}\over\partial
t_{\mu}}dt_{\alpha}, \mu=1, \cdots, r , \ee

\be\label{(z)} dz=(-H_{\alpha} +p_{a}{\partial
H_{\alpha}^{'}\over\partial p_{a}})dt_{\alpha}, \ee where
$z=S(t_{\alpha}, q_{a})$ is the Hamilton-Jacobi function.

One  should notice that although we have started with n
generalized coordinates $q_{i}$ and generalized velocities $\dot
q_{i}$ to pass to canonical formulation we have to treat some
generalized momenta dependent and corresponding generalized
coordinates  as free parameters.Thus , we have a phase space of
lower dimension . But this is not sufficient simply because that
the equations of motion are total differential equations and we
should consider integrability conditions. In other words eqs.
(\ref{{pq}},\ref{(z)} ) are integrable iff $dH_{\alpha}{'}=0$ .
Some of these conditions could be satisfied identically and the
rest may cause new constraints.Again using the same test the
additional constraints, might arise.As a result , it may happen
that we have a set of constraints which are in involution and an
integrable system. Every new constraint causes  to reduce the
dimension of the phase space.
In the end we may have constraints in the form
\be
H_0^{'} =p_0 + H_0,
H_{\gamma}^{'}=H_{\gamma}(t_{\beta}, q_{a}, p_{a}) + p_{\gamma}, \ee
and  additional constraints which can not be expressed in this form.

The equations of motion take the form

\be\label{(q)} dq_{b}={\partial H_{0}^{'}\over\partial
p_{b}}d\tau+ {\partial H_{\gamma}^{'}\over\partial
p_{b}}dt_{\gamma}. \ee

\be\label{(p)} dp_{b}=-{\partial H_{0}^{'}\over\partial
q_{b}}d\tau - {\partial H_{\gamma}^{'}\over\partial
q_{b}}dt_{\gamma}. \ee
Thus, we have an integrable system with some additional
constraints.

The action can be obtained solving  the following equation by quadratures:
\be\label{z1} dz= (-H_{0} + p_{a}{\partial
H_{0}^{'}\over\partial p_{a}})d\tau + (-H_{\beta} + p_{a}{\partial
H_{\beta}^{'}\over\partial p_{a}})dt_{\beta}.\ee

This paper is organized as follows.

In Section 2 the non-relativistic particle moving on a hypersurface in a curved
manifold is investigated using the Hamilton-Jacobi formalism.
The multidimensional rotator is analyzed and  the results are compared
with those obtained by the Dirac Hamiltonian formalism.
In Section 3 the quantization of the multidimensional rotator is investigated.
In Section 4 concluding remarks are presented.

\section{Hamilton-Jacobi formalism of the
non-relativistic particle moving on a hypersurface in a curved
manifold}
We consider a n-dimensional manifold equipped with the  Riemannian metric
$g_{ij}(x)$. Let  $x^{i} (i= 1,2,\cdots n)$ be the coordinates
of the manifold. We consider a non-relativistic particle of mass m whose
motion is constrained on the hypersurface defined as \cite{girotti}
\be
f(x)= B , (B=const).
\ee
In the presence of the vector and scalar potentials $A_i(x)$ and $V(x)$,
the Lagrangian is given by
\be
L={1\over 2}g_{ij}{\dot x^{i}} {\dot x^{j}} + A_i{\dot x^i}- V(x) +
{\dot\lambda} (f(x)-B).
\ee
Here ${\dot x^i}= {dx^i\over dt}$ and $x^{i}(t)$ denotes the position
of the particle , $\lambda$ is a Lagrange multiplier and
${\dot\lambda}= {d\lambda\over dt}$. The canonical momenta conjugate to
$x^i$ and  ${\lambda}$ are
\be
p_{i}= mg_{ij}{\dot x^i} + A_{i},
p_{\lambda}=f(x)- B.
\ee

This Lagrangian leads us to the following  Hamiltonians

\ba\label{fg}
 &H_{0}^{'}&= p_{0} + {1\over 2m}g^{ij}(p_i-A_i)(p_j-A_j)+ V(x) ,\cr
 &H_{1}^{'}&= p_{\lambda}- f(x) + B.\ea

The canonical equations are

\ba\label{hg}
&dx^{i}&= {g^{ij}\over m}(p_{j}- A_{j})dt, \cr
&dp_{i}&= \{{1\over 2m}{\partial g^{lj}\over\partial x^i}(p_l- A_i)(p_j- A_j) +
{\partial V\over\partial x^i} - {g^{lj}\over m}{\partial A_l\over\partial x^i}(p_j-A_j)\}dt
+ {\partial f\over\partial x^i}d\lambda
,\cr
&dp_{\lambda}&=0. \ea

Imposing the variations of   (\ref{fg}) to be zero and take into account
(\ref{hg}) we found immediately the
consistency condition \be\label{ncon} df(x)=0. \ee
Using (\ref{ncon}) a new constraint arises

\be\label{rt}
H_{2}^{'}={g^{ij}\over m}{\partial f\over\partial x^i} (p_{j}-A_{j}).\ee

Taking the variation of (\ref{rt}) and using (\ref{hg}) we obtain
\ba\label{ty}
&{1\over m^{2}}&\{g^{kl}[{\partial\over\partial x^{k}}(g^{ij}
{\partial f\over\partial x^{i}})p_{j} - {\partial\over\partial x^{k}}(g^{ij}
{\partial f\over\partial x^{i}} A_{j})](p_{l}-A_{l}) -
{1\over 2}{\partial g^{kl}\over\partial x^{i}}g^{ij} {\partial f\over\partial x^{j}} p_{k} p_{l}\cr
&+& {\partial\over\partial x^{j}}(g^{kl}A_{l})g^{ij}{\partial f\over\partial x^{i}}p_{k}\}
-{1\over m}g^{ij}{\partial f\over\partial x^{i}}{\partial V\over\partial x^{j}}
+ {g^{ij}\over m}{\partial f\over\partial x^{i}}{\partial f\over\partial x^j}{\dot\lambda}=0.
\ea

Solving (\ref{ty}) we find the Lagrange multiplier
$\lambda$.

In order to compare our results with those obtained using  Dirac's procedure
we analyze the variations of  $H_{0}^{'}, H_{1}^{'},H_{2}^{'}$.
Using
\ba
&dH_{1}^{'}&=\{H_{0}^{'},H_{1}^{'}\}dt,\cr
&dH_{2}^{'}&=\{H_{0}^{'},H_{2}^{'}\}dt + \{H_{0}^{'},H_{1}^{'}\}d\lambda
\ea
we can prove easily that the integrability conditions of (\ref{hg}) are
the same as Dirac's consistency conditions.

\subsection{Multidimensional rotator}

In order to clarify our method we will analyze in detail the multidimensional rotator
problem.
The Lagrangian for a particle of
unit mass constrained to move on the surface of an N-dimensional
sphere of radius R is given by the well known expression
\be\label{lagr}
L={1\over 2}{\dot x_{\alpha}}{\dot
x^{\alpha}}-{\dot\lambda}(x_{\alpha}x^{\alpha} -R^2), \alpha=1\cdots
N, \ee where the constraint
\be
f(x)=-x_{\alpha}x^{\alpha}+R^2=0 \ee is implemented by the
Lagrangian multiplier ${\dot\lambda}$.
Using (\ref{fg}), (\ref{hg}), (\ref{rt}) and (\ref{lagr})
we get a new constraint
\be\label{const}
x^{\alpha} p_{\alpha}=0
\ee
and an equation for $\lambda$ as
\be
{\dot\lambda}={p_{\alpha}p^{\alpha}\over {2x_{\alpha}x^{\alpha}}}.
\ee
Here
$\lambda$ is a gauge parameter.To summarize we have the following set
of Hamiltonians
\be\label{set}
 H_{0}^{'}= p_{0} +{1\over 2 }p_{\alpha}p^{\alpha} ,
 H_{1}^{'}=p_{\lambda}+x_{\alpha}x^{\alpha} - R^2,
 H_{2}^{'}= x^{\alpha} p_{\alpha}.
\ee

The transformation from Cartesian to  curvilinear
coordinates is defined as

\ba
x_{1}&=&r\sin\varphi_{1}\cdots\sin\varphi_{N-1},\quad
x_{2}=r\sin\varphi_{1}\cdots\sin\varphi_{N-2}\cos\varphi_{N-1},\cr
x_{N-3}&=&r\sin\varphi_{1}\sin\varphi_{2}\cos\varphi_{3},\cr
\cdots&=&\cdots,\cr
x_{N-1}&=&r\sin\varphi_{1}\cos\varphi_{2},\quad x_{N}=r\cos\varphi_{1}.
 \ea

In these new variables ,the Lagrangian and the canonical momenta are given
as
\be\label{lagcan}
L={1\over 2}({{\dot r}^2 +r^2{\dot\varphi_{1}}^2 +\cdots r^2{\dot\varphi_{N-1}}^2\cdots\sin^2\varphi_{N-2}})
+{\dot\lambda}(-r+R),
\ee
\ba
&\pi_{\lambda}=-r+R&, \pi_{r}={\dot r},\pi_{\varphi_{1}}=r^2{\dot\varphi_{1}},
\pi_{\varphi_{2}}=r^2{\sin^2\varphi_{1}}{\dot\varphi_{2}},\cr
&\cdots &,\cr
&\pi_{\varphi_{N-1}}=&r^2\sin^2\varphi_{1}\cdots\sin^2\varphi_{N-1}{\dot\varphi_{N-1}}.
\ea
In the Hamilton-Jacobi formalism  we have two Hamiltonians
\ba\label{este}
&H_{0}^{'}&=p_{0}+
{1\over 2 }(\pi_{r}^2 +{\pi_{\varphi_{1}}^2\over r^2\sin^2\varphi_{1}}+ \cdots +
{\pi_{\varphi_{N-1}}^2\over r^2\sin^2\varphi_{1}\cdots r^2\sin^2\varphi_{N-1}}),\cr
&H_{1}^{'}&= \pi_{\lambda} +r-R.
\ea
Using the consistency conditions  $dH_{0}^{'}=0$ and $dH_{1}^{'}=0$
 from  (\ref{hg}) we obtain
 \be\label{com}
 dr=0,d\pi_{r}=0.
 \ee
 Taking into account (\ref{este}) and (\ref{hg}) we have
 \be\label{r}
 d\pi_{r}=-d\lambda + {1\over r^3}( {\pi_{\varphi_{1}}^2\over\sin^2\varphi_{1}}+ \cdots +
{\pi_{\varphi_{N-1}}^2\over\sin^2\varphi_{1}\cdots\sin^2\varphi_{N-1}})dt.
 \ee
 From (\ref{com}) and (\ref{r}) we find
 \be
 {\dot\lambda}= {1\over r^3}(\pi_{r}^2 +{\pi_{\varphi_{1}}^2\over\sin^2\varphi_{1}}+ \cdots +
{\pi_{\varphi_{N-1}}^2\over\sin^2\varphi_{1}\cdots\sin^2\varphi_{N-1}}).
  \ee

In general the physical variables are non-linear functions
of the original variables of the system.
The separation of local coordinates into the physical and pure gauge ones
can be  performed by choosing the curvilinear coordinates in such a way
that some of them span gauge orbits, while the other change
along the directions transverse to the gauge orbits and denote
physical states ( for more details see Refs. \cite{SH,PR}). As an example
we consider the three-dimensional case,
in which the transformation from Cartesian to  spherical coordinates
is given as

\ba &x_1&=r\sin\theta\cos\phi,
x_2=r\sin{\theta}\sin\varphi,\quad x_3=r\cos\theta,\cr
&\pi_{1}&=\sin\theta\cos\varphi\pi_{r}+r\cos\theta\cos\varphi\pi_{\theta}-
r\sin\theta\sin\varphi\pi_{\varphi},\cr
&\pi_{2}&=\sin\theta\sin\varphi\pi_{r}+r\cos\theta\sin\varphi\pi_{\theta}+
r\sin\theta\cos\varphi\pi_{\varphi},\cr
&\pi_{3}&=\cos\theta\pi_{r} -r\sin\theta\pi_{\theta},\cr
& \lambda&=\lambda ,\pi_{\lambda}= p_{\lambda}.  \ea
This transformation is a canonical transformation \cite{caratheo, kastrup}.
 The
canonical pairs are now well defined
$(r,\pi_{r}),(\theta,\pi_{\theta})$ and $(\phi,\pi_{\phi})$.

The Hamiltonians have the following expressions
\ba\label{hamilto}
&H_{0}^{'}&=p_{0}+{\pi_{\theta}^{2}\over 2 r^2} +
{\pi_{\varphi}^2\over 2r^2\sin^2\theta} + {\pi_{r}^{2}\over 2},\cr
&H_{1}^{'}&=\pi_{\lambda} +r-R \ea

and using (\ref{hamilto})
we obtain the following canonical equations
\ba\label{ecuatii}
&dr=&\pi_{r}dt,\quad
d\theta={\pi_{\theta}\over r^2}dt,\quad
d\varphi={\pi_{\varphi}\over r^2\sin^2\theta}dt,\cr
&d\pi_{r}&=-d{\lambda} +{1\over r^3}\left(\pi_{\theta}^2 +
{\pi_{\varphi}^{2}\over\sin^2\theta}\right)dt,\quad d\pi_{\varphi}=0,\cr
&d\pi_{\theta}&={\displaystyle \pi_{\varphi}^2\over r^2\sin^3\theta}\cos\theta
dt,\nonumber\\[5pt]
&d\pi_{\lambda}&=0.
 \ea

Imposing  $dH_{0}^{'}=0$ and $dH_{1}^{'}=0$ we obtain
$H_{2}^{'}=r\pi_r=0$.
Taking into account the consistency condition obtained above and  using  (\ref{com})
we find
\be
{\dot\lambda}={1\over r^3}\left(\pi_{\theta}^2 +
{\pi_{\varphi}^{2}\over\sin^2\theta}\right)
\ee

In this case the action has the following expression
\be
z={1\over 2 R^2}\int\left(\pi_{\theta}^{2}+
{\pi_{\varphi}^2\over\sin^2\theta}\right)dt
\ee

\section{Quantization of the multidimensional\\ rotator}

The multidimensional rotator is a system having second-class constraints
in Dirac's classification  of the constrained systems. The Hamilton -Jacobi
formalism  leads us to three Hamiltonians  $H_{0}^{'}, H_{1}^{'},H_{2}^{'}$
which are not in involution. At this stage we mention that it is possible
always to make the Hamltonians in involution  and then the corresponding
new system is integrable.
In our specific problem   we can apply
the method of Abelian conversion to transform the system
into an Abelian gauge theory \cite{bata}.

We found the Hamiltonians in involution :
\be\label{cuantic}
 H_{0}^{''}= p_{0} + {1\over 2}\left({{(H_{2}^{"})^2\over H_{1}^{"} +R^2} +
           {L_{a}^2\over H_{1}^{''} +R^2}}\right)
, H_{1}^{''}=p_{\lambda}+x_{\alpha}x^{\alpha} - R^2,
 H_{2}^{''}= x^{\alpha} p_{\alpha} +2x^2\lambda,
 \ee
 where $L_{a}=-ipL_{a}x$ is the classical component of the angular momentum
 (with $a =i,j$ , $L_{a}=x_{i}p_{j}-x_{j}p_{i}$).

At the quantum level we obtain  $ H_{0}^{''}\Psi= H_{2}^{''}\Psi=
H_{1}^{''}\Psi= 0$,
 where $\Psi$ is the wave function.
 The first-class constraints restrict the physical Hilbert space to the
 gauge-invariant sector

\be\label{cuantic1}
 H_{1}^{''}\Psi_{phys}=0, \quad H_{2}^{''}\Psi_{phys}=0.
\ee
The general solution of (\ref{cuantic1}) has the following form:
\be
\Psi_{phys}=f(\lambda, x^2)\Psi(\Omega),
\ee
where $f(x,\lambda)$ is some function , whereas $\Psi(\Omega)$
is wave function on the N-sphere.
 In the physical Hilbert space, we make
$H_{1}^{''},H_{2}^{''}$
 zero in $H_{0}^{''}$.
Taking into account (\ref{cuantic}) and (\ref{cuantic1})
we immediately find the energy values
\be
E_{l}={{\hbar}^2\over 2R^2}l(l+N-1)
\ee
 and conclude that the quantum
Hamiltonian for the multidimensional rotor is given by the pure
Schr\"{o}dinger operator without any boundary term.

\section{Concluding remarks}

 Despite the success of Dirac's approach in studying singular systems, which is
 demonstrated by the wide number of physical systems to which this
 formalism has been applied, it is always instructive to study singular systems
 through other formalisms, since different procedures will provide different
 views for same problems , even for non-singular systems.

 In the Hamilton-Jacobi formalism  we have a set of partial
differential equations to start with, and we construct the phase
space using the integrability conditions of a set of total
differential equations.
In this formalism we have no distinction between the  first- and the
second- class constraints  but the  Dirac's consistency
conditions are equivalent to the Hamilton-Jacobi integrability conditions.

In this paper the Hamilton-Jacobi formalism was applied to investigate the
non-relativistic particle moving on a hypersurface in a curved manifold
and we found the same set of constraints as by using  Dirac's approach.
In the case of the multidimensional rotator we have eliminated the
non-physical degrees of freedom transforming the Cartesian coordinates
into curvilinear ones.For the three- dimensional rotator a canonical
transformation was performed in order to find the physical degrees of
freedom and  the action was calculated.

As pointed in \cite{kleinert} the energy spectrum
of the multidimensional rotator obtained
by  Dirac's quantization method must be rejected because
it is physically incorrect.
Using the fact that $\lambda$ is a gauge parameter
in the Hamilton-Jacobi formalism  we found
the same result as in \cite{kleinert}.
The quantum Hamiltonian of the multidimensional rotor is given by pure
the Schr\"{o}dinger operator without any boundary term.

As a further step we will apply  this method to non-Abelian gauge
theory and gravity. This programme is under investigation, and this
 article is the first step in this direction.

\section {Acknowledgements}
DB would like to thank M. Olshanetsky for useful
discussions.This paper is partially supported by the
Scientific and Technical Research  Council of Turkey.

\end{document}